\documentclass{article}
\usepackage{spconf,amsmath,graphicx, booktabs}
\usepackage[hidelinks]{hyperref}
\usepackage{amssymb}
\usepackage{bm}

\title{Point2Acoustic: Efficient Room Acoustic Parameter Estimation from 3D Geometry and Material Properties}
\name{
Xun Gao$^{1,2}$ \qquad
Raksha Ramakrishna$^{2}$ \qquad
Tommy Falk$^{2}$ \qquad
Werner de Bruijn$^{2}$
}

\address{
$^{1}$KTH Royal Institute of Technology, Stockholm, Sweden \qquad
$^{2}$Ericsson Research, Stockholm, Sweden
}
\begin{document}
\ninept
\maketitle

\begin{abstract}
Scene-adaptive immersive audio rendering requires rapid estimation of acoustic parameters that reflect the surrounding environment. For repeated runtime scene updates, performing a detailed geometric-acoustic simulation for every update can be computationally costly. We present Point2Acoustic, a learning-based framework for direct estimation of six-band reverberation time ($RT_{60}$) and reverberant-to-direct ratio (RDR) from 3D scene geometry and acoustic material properties. Each scene is represented by a surface-sampled point cloud with spatial coordinates and frequency-dependent absorption and scattering coefficients, and encoded using PointNet or PointNet++. We train and evaluate the framework on a synthetic furnished-room dataset with varied geometry and acoustic materials, using geometric-acoustic simulation to obtain reference targets. Point2Acoustic achieves accurate multi-band prediction with millisecond-scale inference, and ablation results indicate that explicit global room descriptors and semantic surface labels provide no consistent benefit. The proposed formulation supports efficient scene-adaptive acoustic rendering in extended reality (XR).
\end{abstract}


\section{INTRODUCTION}
\label{sec:introduction}

Realistic spatial audio is an important component of immersive extended-reality (XR) systems, where the rendered sound should adapt to changes in the surrounding physical environment. Scene-adaptive spatial-audio rendering, as supported by MPEG-I Immersive Audio~\cite{iso23090_4}, requires efficient estimation of room-acoustic parameters that characterize the reverberant environment. In this work, we focus on frequency-dependent reverberation time ($RT_{60}$) and reverberant-to-direct ratio (RDR), which provide compact control parameters for late-reverberation rendering without requiring the complete RIR. These parameters are commonly estimated from measured or simulated room impulse responses (RIRs), but RIRs can be cumbersome to measure and computationally expensive to simulate for frequent scene updates. This motivates a direct scene-to-acoustic mapping that can be learned offline and rapidly inferred at deployment.

Modern XR devices can reconstruct geometric information about the surrounding environment using scene-understanding techniques~\cite{meta_scene}. Point-based representations can directly encode reconstructed scene geometry and material properties, avoiding the need to derive handcrafted global room descriptors. Acoustic behavior depends not only on geometry but also on the spatial distribution of surface materials. We represent each scene using 3D surface samples augmented with frequency-dependent material absorption and scattering coefficients. This representation preserves both geometric structure and local acoustic material information, in contrast to global descriptors such as room dimensions, volume, or scene-averaged absorption.

Room-acoustic properties are traditionally estimated using analytical models from statistical room acoustics, such as Sabine and Eyring, or through geometrical and wave-based simulation~\cite{vorlander2020auralization}. Learning-based approaches have also been investigated for predicting acoustic parameters from audio, compact geometric descriptors, and spatial representations of room geometry and absorption~\cite{groger2025evaluation, falcon2021spherical}. Related neural acoustic modeling has focused on RIR synthesis and spatial acoustic-field reconstruction using room descriptors, triangular meshes, implicit neural fields, or sparse acoustic observations combined with scene geometry~\cite{ratnarajah2022fast,ratnarajah2022mesh2ir,luo2022learning,su2022inras,wang2024hearing}. Point-based representations have also been explored in acoustic modeling~\cite{meng2021pointbased,lluis2022points2sound}, but these methods do not directly estimate compact room-acoustic parameters. Many of these methods target position-dependent acoustic responses or full room impulse responses. Our goal is a compact scene-level parameterization that does not require source or receiver coordinates as inputs. Direct prediction of compact multi-band room-acoustic parameters from 3D geometry together with spatially distributed acoustic material properties remains less explored.

To address this problem, we propose Point2Acoustic, a point-cloud regression framework that directly maps a 3D scene representation to six-band $RT_{60}$ and RDR. PointNet~\cite{qi2017pointnet} and PointNet++~\cite{qi2017pointnetpp} are evaluated as alternative geometric encoders within the same formulation. Rather than relying on explicitly derived room dimensions, volume, surface areas, or globally averaged material statistics, Point2Acoustic learns directly from spatial samples carrying local acoustic properties. At deployment, the learned mapping replaces repeated geometric-acoustic simulation with a single network inference.

\textbf{The main contributions of this work are:} (1) We propose Point2Acoustic, a direct scene-to-acoustic formulation that estimates six-band $RT_{60}$ and RDR from 3D geometry and frequency-dependent acoustic material properties, replacing repeated geometric-acoustic simulation at deployment with a single network inference. (2) To the best of our knowledge, we are the first to estimate scene-level acoustic parameters directly from 3D point clouds with per-point absorption and scattering coefficients. (3) We construct a synthetic furnished-room dataset with diverse geometry and acoustic materials and evaluate Point2Acoustic against analytical and learned baselines in terms of prediction accuracy, point-cloud robustness, and computational efficiency.

\section{POINT2ACOUSTIC: 3D SCENE-TO-ACOUSTIC PREDICTION}
\label{sec:model}

\subsection{Acoustic Scene Representation}

We formulate room-acoustic parameter estimation as a set-to-vector regression problem. Let $B=6$ denote the octave bands considered in this work, spanning 125~Hz to 4~kHz. A scene is represented by a surface-sampled point set
\begin{equation}
\mathcal{P}=\{\mathbf{x}_j\}_{j=1}^{N}, \qquad
\mathbf{x}_j =
\left[
\mathbf{p}_j^{\mathsf T},
\bm{\alpha}_j^{\mathsf T},
\bm{\sigma}_j^{\mathsf T}
\right]^{\mathsf T}
\in\mathbb{R}^{3+2B},
\end{equation}
where $\mathbf{p}_j\in\mathbb{R}^{3}$ is the Cartesian coordinate of point $j$, and $\bm{\alpha}_j,\bm{\sigma}_j\in\mathbb{R}^{B}$ are the frequency-dependent absorption and scattering coefficients of the corresponding surface. The superscript $(\cdot)^{\mathsf T}$ denotes transpose. 

Points are sampled from the exposed room and furniture surfaces and inherit their corresponding acoustic material properties. We assume that acoustic material properties are available for the input scene; their estimation is not considered in this work. 

\subsection{Acoustic Targets and Scene-to-Acoustic Regression}

Point2Acoustic predicts six-band $RT_{60}$ and RDR, with the scene-level target defined as
\begin{equation}
\mathbf{y}
=
[\mathbf{RT}_{60}^{\mathsf T},
\mathbf{RDR}^{\mathsf T}]^{\mathsf T}
\in\mathbb{R}^{2B},
\qquad
\mathbf{RT}_{60},\mathbf{RDR}\in\mathbb{R}^{B}.
\label{eq:acoustic_target}
\end{equation}
$RT_{60}$ characterizes the reverberation decay time and is derived from the Schroeder energy-decay curve of the band-limited RIR~\cite{schroeder1965}. RDR characterizes the relative strength of reverberation in a space. The conventional direct-to-reverberant ratio (DRR) varies with source--receiver distance and source directivity. RDR instead references the late-reverberant and direct-sound energies to a 1~m distance from an omnidirectional source, making it a scene-level measure of the acoustic environment. Following the MPEG-I definition~\cite{iso23090_4} and its Listener Space Information (LSI) reference implementation~\cite{iso23090_34}, RDR for frequency band $b$ is defined as
\begin{equation}
\mathrm{RDR}_b
=
10\log_{10}
\left(
\frac{
\displaystyle\int_{t_2}^{t_{\mathrm{end},b}} s_b^2(t)\,dt
}{
\displaystyle r_{\mathrm{meas}}^2
\int_{t_{0,b}}^{t_{1,b}} s_b^2(t)\,dt
}
\right)\,\mathrm{dB},
\label{eq:rdr}
\end{equation}
where $s_b(t)$ is the band-limited RIR, $r_{\mathrm{meas}}$ is the source--receiver distance, $t_{0,b}$ and $t_{1,b}$ delimit the interval that contains the bulk of the direct-sound energy, and $t_2$ and $t_{\mathrm{end},b}$ delimit the late-reverberation interval. The factor $r_{\mathrm{meas}}^2$ compensates for the distance dependence of the direct-path energy, making the resulting RDR independent of the measurement distance. The diffuse-field onset $t_2$ is estimated from the broadband normalized echo-density profile according to MPEG-I~\cite{iso23090_4} and shared across frequency bands.

Point2Acoustic learns the scene-to-acoustic mapping
\begin{equation}
f_{\Theta}:\mathcal{P}\rightarrow\mathbb{R}^{2B},
\qquad
\hat{\mathbf{y}}
=
f_{\Theta}(\mathcal{P})
=
[\widehat{\mathbf{RT}}_{60}^{\mathsf T},
\widehat{\mathbf{RDR}}^{\mathsf T}]^{\mathsf T}.
\label{eq:scene_mapping}
\end{equation}
The mapping is decomposed into a geometric encoder and task-specific regression heads,
\begin{equation}
\mathbf{z}=E_{\theta}(\mathcal{P}), \qquad
\hat{\mathbf{y}}=H_{\phi}(\mathbf{z}),
\label{eq:model_decomposition}
\end{equation}
where $\mathbf{z}$ is a shared global scene representation and $H_{\phi}$ comprises two multilayer-perceptron heads for $RT_{60}$ and RDR, with $\Theta=\{\theta,\phi\}$.

\begin{figure*}[t]
    \centering
    \includegraphics[
        width=\textwidth,
        trim={0cm 5.3cm 2.7cm 2.4cm},
        clip
    ]{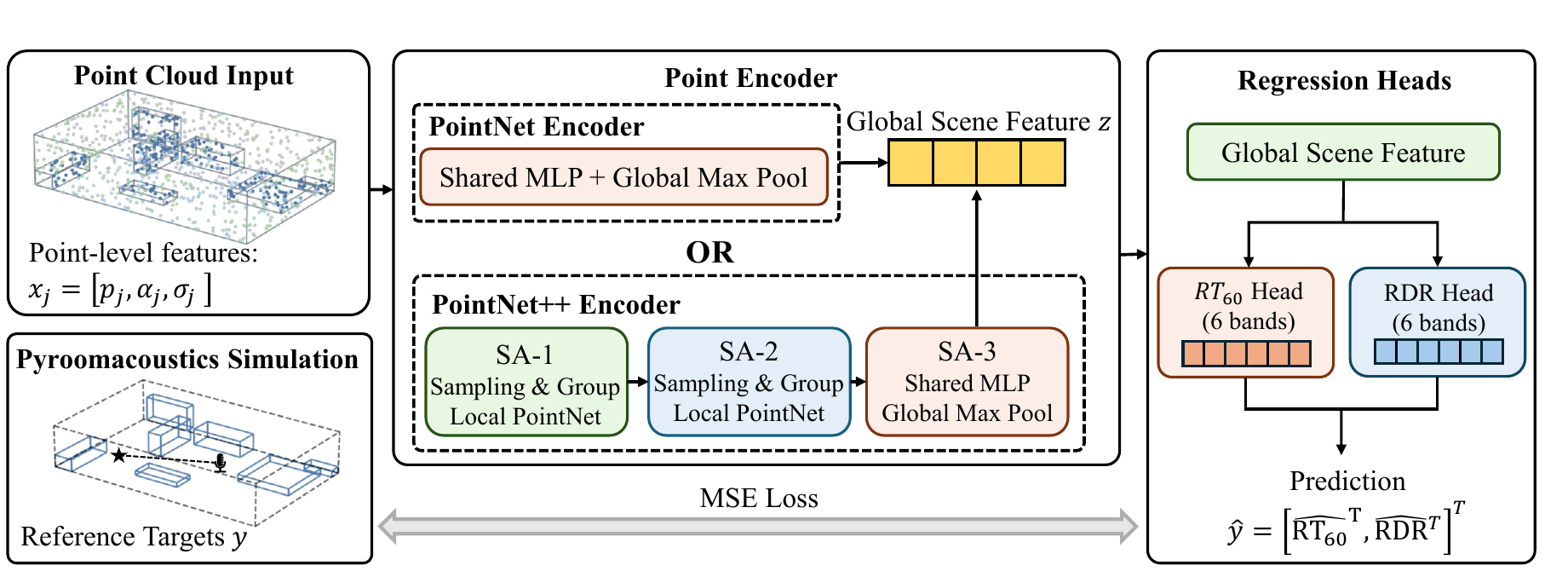}
    \caption{Overview of the Point2Acoustic training framework. Surface-sampled 3D scenes are encoded by PointNet or PointNet++ and mapped to six-band $RT_{60}$ and RDR estimates using simulation-derived training targets.}
    \label{fig:model}
\end{figure*}

We investigate two variants of Point2Acoustic based on different geometric encoders $E_{\theta}$. Point2Acoustic-PN uses PointNet~\cite{qi2017pointnet}, which produces a global permutation-invariant scene representation through shared point-wise transformations and symmetric pooling. Point2Acoustic-PN++ uses PointNet++~\cite{qi2017pointnetpp}, which additionally models local geometric structure through hierarchical set abstraction. Their comparison evaluates the benefit of local geometric modeling for scene-level acoustic prediction. Figure~\ref{fig:model} summarizes the training framework.

\subsection{Training Objective}

To balance the two acoustic targets across frequency bands, each target dimension is standardized using training-set statistics,
\begin{equation}
\tilde{y}_{k,b}
=
\frac{y_{k,b}-\mu_{k,b}}{s_{k,b}},
\qquad
k\in\{RT_{60},\mathrm{RDR}\},
\quad b=1,\ldots,B,
\label{eq:standardization}
\end{equation}
where $\mu_{k,b}$ and $s_{k,b}$ denote the training-set mean and standard deviation for target $k$ and frequency band $b$. The network is optimized jointly over the two acoustic targets using
\begin{equation}
\mathcal{L}(\Theta)
=
\frac{1}{B}
\sum_{b=1}^{B}
\sum_{k}
w_k
\left(
\hat{\tilde{y}}_{k,b}
-
\tilde{y}_{k,b}
\right)^2,
\label{eq:training_loss}
\end{equation}
where $w_k$ controls the relative contribution of each task. We use equal weighting, $w_{RT_{60}}=w_{\mathrm{RDR}}=1$.

\section{SYNTHETIC FURNISHED-ROOM DATASET}
\label{sec:dataset}

We construct 5,000 randomized furnished single-room scenes with varying room dimensions, object layouts, and frequency-dependent acoustic materials. Horizontal dimensions are sampled uniformly from 3--12~m, with heights between 2.4 and 4.0~m. Single-room environments provide a controlled setting and avoid the multi-slope decay behavior of acoustically coupled spaces~\cite{bradley2009quantifying}. Shoebox rooms are populated with axis-aligned cuboid furniture to introduce geometric variability with simple scene generation.

Furniture is placed near walls, in corners, or in free interior regions, with 2--10 objects per room subject to boundary-clearance and non-overlap constraints. Materials are sampled from category-dependent pools for walls, floors, ceilings, and furniture. The database contains 19 materials with six-band absorption coefficients of 0.01--0.88 and scattering coefficients of 0.05--0.65. Absorption data are taken from the Pyroomacoustics database~\cite{vorlander2020auralization,pyroomacoustics}; scattering profiles are assigned from qualitative surface classes following practical room-acoustic modeling recommendations~\cite{zeng2006scattering} and are treated as approximate simulation parameters.

Reference RIRs are simulated with Pyroomacoustics~\cite{pyroomacoustics} using the room and exposed furniture surfaces with their assigned materials. For each scene, 10 source–receiver pairs are sampled, with both positions in free space at least 0.25~m from boundaries and objects and with direct line of sight between them. Scene-level $RT_{60}$ and RDR targets are taken as the median across the 10 pairs to reduce sensitivity to atypical source--receiver configurations.

\section{EXPERIMENTS AND RESULTS}
\label{sec:experiments_and_results}

\subsection{Experimental Setup}
\label{subsec:experimental_setup}

The 5,000 scenes are split at the scene level into 4,000 training, 500 validation, and 500 test scenes. Each scene is represented by $N=8192$ surface samples, using a fixed point budget for consistent network input size and computational cost across scenes. Samples are distributed proportionally to surface area. All neural models are trained using AdamW~\cite{loshchilov2019decoupled} with an initial learning rate of $5\times10^{-4}$ and batch size 32. Performance is evaluated using mean absolute error (MAE) and the coefficient of determination $R^2$. For the computational-efficiency evaluation, Pyroomacoustics runtime is reported as mean $\pm$ standard deviation over 50 scenes using a single CPU core, and neural inference is reported per scene with batch size 32 on an NVIDIA RTX PRO 6000 GPU. Additional experiments examine feature importance and robustness to point-cloud perturbations.

\subsection{Acoustic Parameter Prediction}
\label{subsec:prediction}

Estimates from the statistical Sabine and Eyring equations are included as analytical baselines for $RT_{60}$ using room volume, surface area, and frequency-dependent absorption~\cite{sabine1922collected,eyring1930reverberation}. Corresponding RDR estimates are obtained using a statistical approximation based on the MPEG-I Immersive Audio formulation~\cite{iso23090_4}. For octave band $b$,
\begin{equation}
\widehat{\mathrm{RDR}}_{b}^{(m)}
=
10\log_{10}
\left(
\frac{RT_{60,b}^{(m)}}{V}
\right)
+25
-
\frac{60t_2}{RT_{60,b}^{(m)}} ,
\label{eq:analytical_rdr}
\end{equation}
where $m\in\{\mathrm{Sab},\mathrm{Eyr}\}$ denotes the analytical model, $RT_{60,b}^{(m)}$ is the corresponding reverberation-time estimate, $V$ is the room volume, and $t_2$ is the onset time of late reverberation used in the RDR definition in~\eqref{eq:rdr}.

A multilayer perceptron (MLP) is included as a learned baseline that predicts the acoustic parameters from global scene descriptors
\begin{equation}
\mathbf{x}_{\mathrm{MLP}} =
[L_x,L_y,L_z,V,
\bar{\boldsymbol{\alpha}},
\bar{\boldsymbol{\sigma}}],
\label{eq:mlp_input}
\end{equation}
where $L_x$, $L_y$, and $L_z$ are the room dimensions, $\bar{\boldsymbol{\alpha}}$ and $\bar{\boldsymbol{\sigma}}$ are six-band surface-area-weighted average absorption and scattering coefficients. The MLP and analytical baselines rely on explicitly derived global room statistics. Point2Acoustic instead operates directly on 3D point samples with per-point acoustic material properties.

Table~\ref{tab:comparison} compares prediction accuracy and computational efficiency across the evaluated methods. Metrics are averaged over the six octave bands on the held-out test set.
\begin{table}[t]
\centering
\caption{Prediction accuracy and computational efficiency of the evaluated methods. PN and PN++ denote Point2Acoustic-PN and Point2Acoustic-PN++, respectively.}
\label{tab:comparison}

\fontsize{9}{9.2}\selectfont
\setlength{\tabcolsep}{0.9pt}
\renewcommand{\arraystretch}{0.9}

\begin{tabular}{
@{}l
@{\hspace{10pt}}
cc
@{\hspace{8pt}}
cc
@{\hspace{6pt}}
c
@{}
c@{}
}
\toprule
& \multicolumn{2}{c}{$RT_{60}$}
& \multicolumn{2}{c}{RDR}
& & \\
\cmidrule(lr){2-3}
\cmidrule(lr){4-5}

\textbf{Method}
& \textbf{MAE $\downarrow$}
& \textbf{$R^2$ $\uparrow$}
& \textbf{MAE $\downarrow$}
& \textbf{$R^2$ $\uparrow$}
& \textbf{Params.}
& \textbf{Runtime} \\

& \textbf{(s)}
&
& \textbf{(dB)}
&
& \textbf{(M)}
& \textbf{/scene} \\

\midrule

Sabine
& 0.1511
& 0.8942
& 1.4645
& 0.6798
& --
& -- \\

Eyring
& 0.1542
& 0.8873
& 1.5557
& 0.6956
& --
& -- \\

\midrule

MLP
& 0.1316
& 0.9299
& 1.1295
& 0.8688
& \textbf{0.020}
& \textbf{0.0022 ms} \\

PN
& 0.1118
& 0.9436
& \textbf{1.1145}
& 0.8717
& 0.473
& 0.927 ms \\

PN++
& \textbf{0.0916}
& \textbf{0.9661}
& 1.1210
& \textbf{0.8733}
& 0.615
& 11.29 ms \\

\midrule

\multicolumn{6}{l}{Pyroomacoustics}
& $31.3\pm16.0$ s \\

\bottomrule
\end{tabular}
\end{table}

\begin{figure}[t]
    \centering
    \includegraphics[width=\columnwidth]{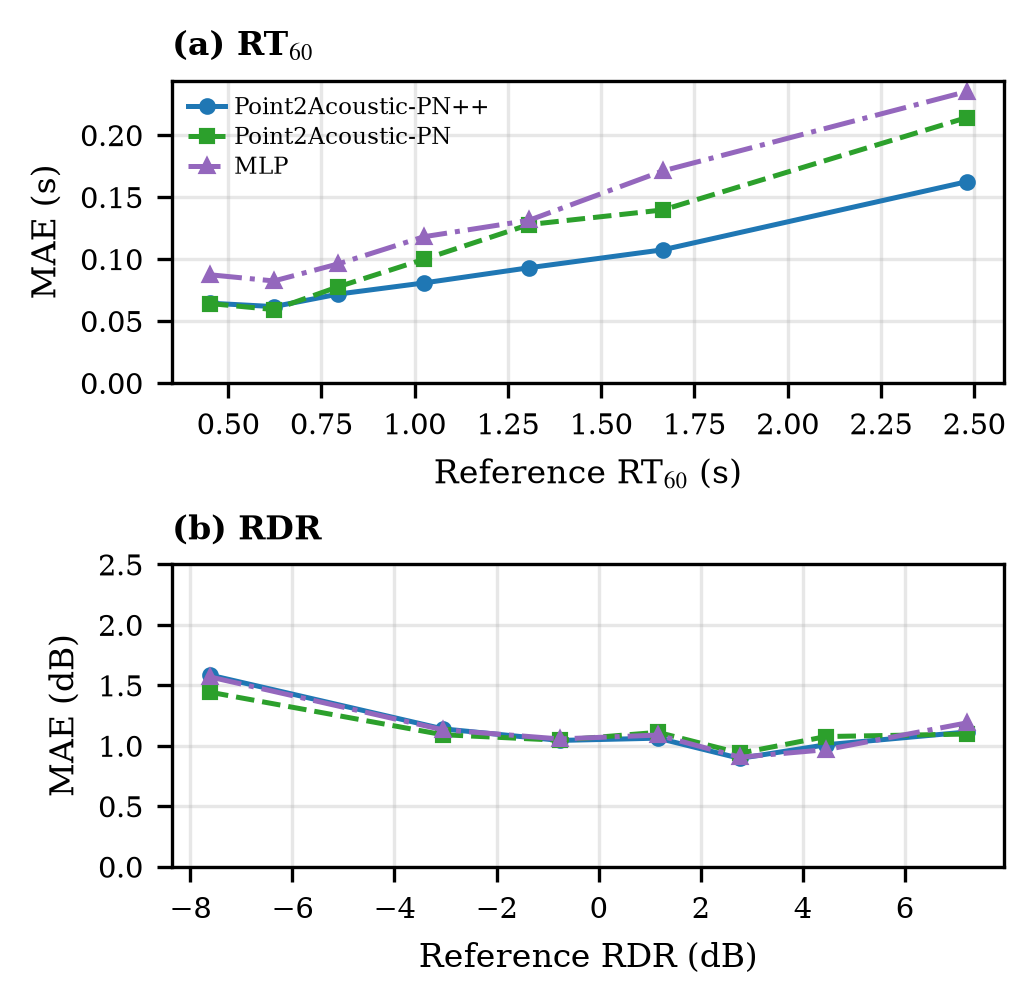}
    \caption{Prediction error across the reference acoustic-parameter range.
    MAE is computed within seven equal-count bins over the six octave bands
    and 500 test scenes.}
    \label{fig:error_vs_reference}
\end{figure}

To place the aggregate errors in Table~\ref{tab:comparison} in the context of
the reference-label values, Fig.~\ref{fig:error_vs_reference} examines how the
prediction error varies across the evaluated acoustic-parameter range. The
simulation-derived reference values are pooled over the six octave bands on
the test set and divided into seven equal-count bins. For target
$k\in\{RT_{60},\mathrm{RDR}\}$, model $m$, and bin $n$, the binned MAE and
corresponding mean reference value are
\begin{equation}
\begin{aligned}
\mathrm{MAE}_{n}^{(k,m)}
&=
\frac{1}{|\mathcal{I}_{n}^{(k)}|}
\sum_{(i,b)\in\mathcal{I}_{n}^{(k)}}
\left|\hat{y}_{i,k,b}^{(m)}-y_{i,k,b}\right|,\\
\bar{y}_{n}^{(k)}
&=
\frac{1}{|\mathcal{I}_{n}^{(k)}|}
\sum_{(i,b)\in\mathcal{I}_{n}^{(k)}}
y_{i,k,b}.
\end{aligned}
\label{eq:binned_mae}
\end{equation}
Here, $\mathcal{I}_{n}^{(k)}$ denotes the set of scene--band samples assigned
to the $n$th bin for target $k$, and $\bar{y}_{n}^{(k)}$ gives the horizontal
coordinate of the corresponding point in Fig.~\ref{fig:error_vs_reference}.

The MLP performs strongly in Table~\ref{tab:comparison}, indicating that much of the acoustic variation in the synthetic dataset can be captured from global room and material statistics. Both Point2Acoustic variants further improve $RT_{60}$ prediction, with Point2Acoustic-PN++ achieving the best performance. Figure~\ref{fig:error_vs_reference}(a) further shows that its advantage becomes more pronounced toward larger $RT_{60}$ values. In contrast, RDR errors remain relatively stable across the evaluated range, and the three learned models exhibit similar performance in Fig.~\ref{fig:error_vs_reference}(b). This suggests limited benefit from hierarchical local geometric modeling for RDR under clean synthetic conditions. Combined with the robustness results in Sec.~\ref{subsec:robustness}, these results show that point-based scene representations can improve prediction without requiring explicitly derived global room statistics.

\subsection{Feature Ablation}

Semantic surface labels and room-level geometric descriptors are evaluated only in the feature-ablation study in Table~\ref{tab:feature_ablation}. ``Full'' uses all features. ``XYZ + material'' uses only XYZ coordinates and per-point absorption and scattering coefficients.

\begin{table}[t]
\centering
\caption{Feature ablation of Point2Acoustic-PN++. Material denotes per-point six-band absorption and scattering coefficients.}
\label{tab:feature_ablation}

\fontsize{9}{9.2}\selectfont
\setlength{\tabcolsep}{1.2pt}
\renewcommand{\arraystretch}{0.9}

\begin{tabular}{@{}lcccc@{}}
\toprule
& \multicolumn{2}{c}{$RT_{60}$}
& \multicolumn{2}{c}{RDR} \\
\cmidrule(lr){2-3}
\cmidrule(lr){4-5}

\textbf{Features}
& \textbf{MAE (s) $\downarrow$}
& \textbf{$R^2$ $\uparrow$}
& \textbf{MAE (dB) $\downarrow$}
& \textbf{$R^2$ $\uparrow$} \\

\midrule

Full
& 0.0979 & 0.9612 & 1.1395 & 0.8708 \\

No global
& 0.0956 & 0.9628 & \textbf{1.0986} & \textbf{0.8766} \\

No semantic
& 0.0941 & 0.9652 & 1.1194 & 0.8718 \\

XYZ + material
& \textbf{0.0916} & \textbf{0.9661} & 1.1210 & 0.8733 \\

No material
& 0.4707 & 0.1532 & 3.3294 & 0.2334 \\

XYZ only
& 0.4756 & 0.1611 & 3.3295 & 0.2271 \\

\bottomrule
\end{tabular}
\end{table}

Material information is the dominant input feature. Removing absorption and scattering coefficients reduces $R^2$ from approximately 0.97 to 0.16 for $RT_{60}$ and from 0.87 to 0.23 for RDR. Adding explicit global room descriptors or semantic surface labels provides no consistent performance gain: the simpler XYZ + material configuration achieves the best $RT_{60}$ results and comparable RDR performance. Based on these results, we retain XYZ coordinates and per-point absorption and scattering coefficients as the final Point2Acoustic input, reducing feature dimensionality and avoiding additional scene-level preprocessing.

\subsection{Robustness to Point-Cloud Perturbations}
\label{subsec:robustness}

\begin{figure}[t]
    \centering
    \includegraphics[width=\columnwidth]{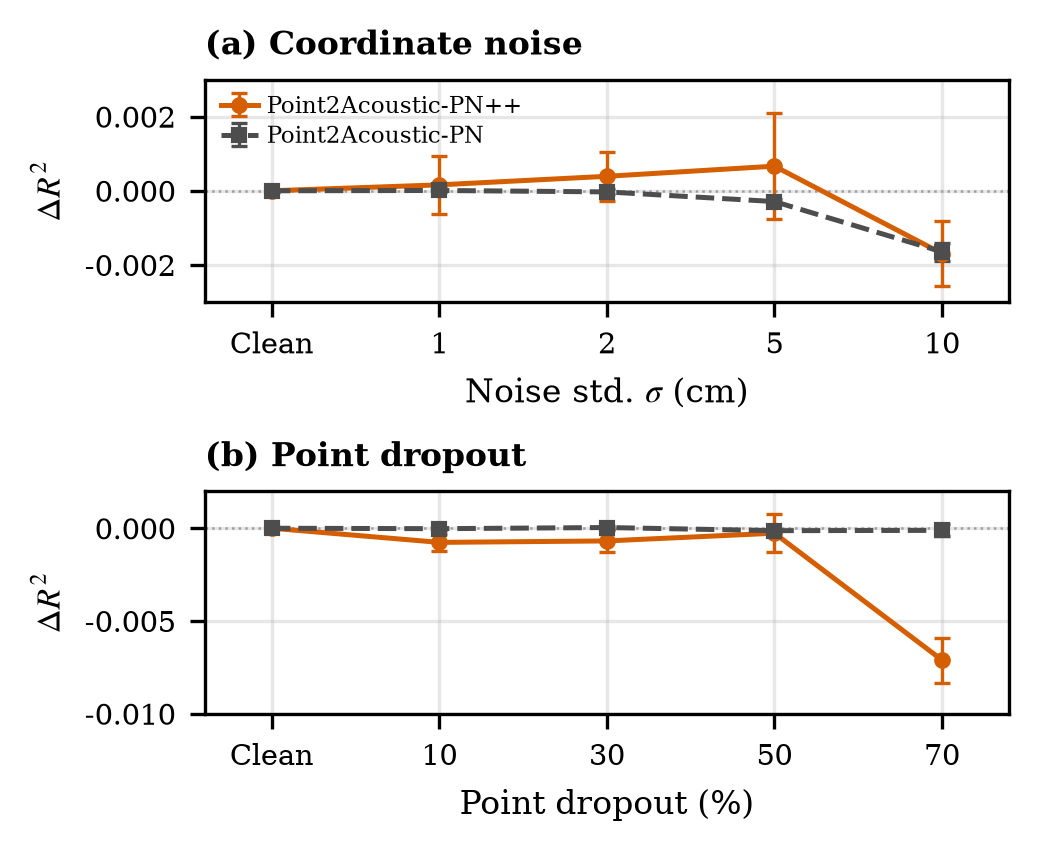}
    \caption{Robustness to coordinate noise and point dropout. Results show the change in test $R^2$ from clean input; error bars denote mean $\pm$ standard deviation over five trials.}
    \label{fig:robustness}
\end{figure}

Figure~\ref{fig:robustness} evaluates sensitivity to Gaussian coordinate noise and point dropout applied to the test set. Robustness is measured relative to clean-input performance as 
\begin{equation}
\Delta R^2=R^2_{\mathrm{perturbed}}-R^2_{\mathrm{clean}}, 
\label{eq:r2}
\end{equation}
where values close to zero indicate little degradation. For point dropout, retained points are resampled with replacement to maintain a fixed input size of \(N=8192\). Performance remains stable under moderate perturbations and degrades gradually at stronger corruption levels. This suggests that Point2Acoustic does not require exact point-level reconstruction and can tolerate moderate noise or point loss in the input geometry.

\subsection{Computational Efficiency}

Detailed geometric-acoustic simulation can require substantial computation for each new scene. Point2Acoustic replaces repeated simulation at deployment with a single network inference. As shown in Table~\ref{tab:comparison}, the Pyroomacoustics-based labeling pipeline requires tens of seconds per scene, compared with millisecond-scale inference for both Point2Acoustic variants. Point2Acoustic-PN++ provides the highest $RT_{60}$ accuracy at 11.29~ms per scene. Point2Acoustic-PN retains most of the predictive performance with sub-millisecond inference, providing a more favorable accuracy--efficiency trade-off for resource-constrained deployment. Although the MLP is faster, both point-cloud models are more accurate and remain sufficiently efficient for integration into real-time XR acoustic-rendering pipelines.

\section{CONCLUSION AND FUTURE WORK}
\label{sec:conclusion}

This work shows that 3D point clouds can serve as a basis for efficiently predicting frequency-dependent $RT_{60}$ and RDR, with no observed benefit from adding explicit global room descriptors. Under clean synthetic conditions, Point2Acoustic-PN++ provides the strongest $RT_{60}$ accuracy, but Point2Acoustic-PN retains comparable RDR performance with substantially lower inference cost. The robustness results also suggest that the method can tolerate coordinate noise and point loss in imperfect point-cloud observations. Future work will extend the evaluation to more complex and irregular scene geometries, where the point-cloud representation is expected to provide greater benefits over global-descriptor MLPs and statistical models. Further work will also include validation in measured environments and integration into an end-to-end XR spatial-audio pipeline.

\vfill\pagebreak

\bibliographystyle{IEEEbib}
\bibliography{refs}

\end{document}